\documentclass{pasj00}
\begin{document}
\SetRunningHead{M.\,Sawada}{Suzaku X-ray Observations of W\,28}
\Received{2011/11/29}
\Accepted{2012/02/04}
\title{X-Ray Observations of the Supernova Remnant W\,28 with Suzaku --- \\ 
I. Spectral Study of the Recombining Plasma}
\author{
Makoto \textsc{Sawada}\altaffilmark{} and
Katsuji \textsc{Koyama}\altaffilmark{}
}
\altaffiltext{}{Department of Physics, Graduate School of Science, Kyoto University, 
Kitashirakawa Oiwake-cho, Sakyo-ku, Kyoto 606-8502}
\email{sawada@cr.scphys.kyoto-u.ac.jp}
\maketitle
\begin{abstract}
We present the Suzaku results of the mixed-morphology supernova remnant W\,28. 
The X-ray spectra of the central region of W\,28 exhibit many bright emission lines from highly 
ionized atoms. An optically thin thermal plasma in  collisional ionization equilibrium, either 
of single-temperature or multi-temperature  failed to reproduce the data with line-like and bump-like 
residuals at the Si Lyman$\alpha$ energy and at 2.4--5.0~keV, respectively.
The bumps probably correspond to radiative recombination continua from He-like Si and S. 
A simple recombining plasma model nicely fit the bump structures, but failed to fit low energy bands. 
The overall spectra can be fit with a multi-ionization temperature plasma with a common 
electron temperature. The multi-ionization temperatures are interpreted as elemental 
difference of ionization and recombination timescales. These results prefer the rarefaction scenario 
for the origin of the recombining plasma.

\end{abstract}
\begin{table*}[!htbc]
 \caption{Log of Suzaku observations of W\,28.}\label{tab:obslog}
 \begin{center}
  \begin{tabular}{cccccc}
   \hline
   Sequence no. & \multicolumn{2}{c}{Aim point} & Start date & Effective & Field name \\
                & $\alpha$ (J2000.0) & $\delta$ (J2000.0) & & exposure & \\ \hline
   505005010 & \timeform{18h00m17s} & \timeform{-23D21'59"} & 2010/04/03 & 73.0 ks & Center \\
   500008010 & \timeform{18h03m49s} & \timeform{-22D01'03"} & 2006/04/07 & 40.7 ks & Nearby sky \\
   \hline
  \end{tabular}
 \end{center}
\end{table*}
\begin{figure*}[!htbc]
  \begin{center}
    \FigureFile(140mm,90mm){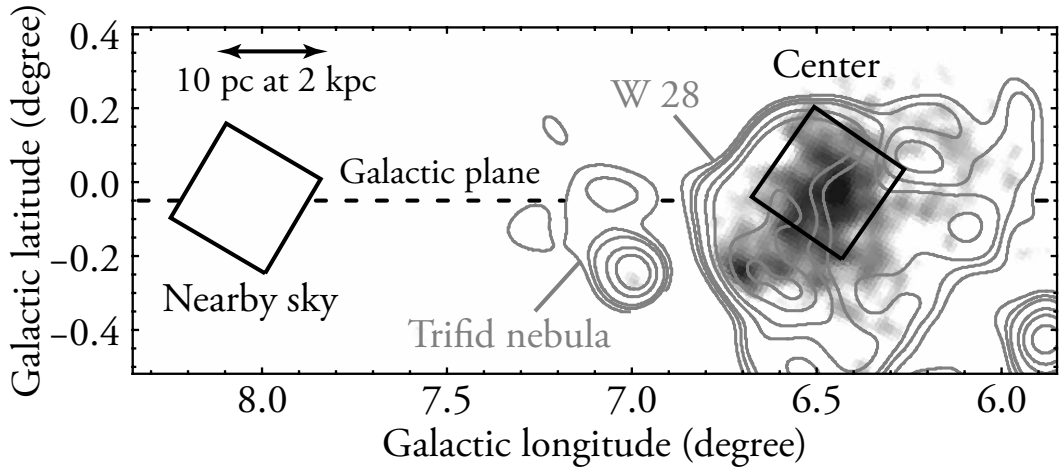}
  \end{center}
  \caption{Wide-field X-ray image retrieved from the ROSAT All Sky Survey in grayscale
(\cite{1995ApJ...454..643S}). Overlaid contours are the radio continuum map at 1.4~GHz by the Very Large Array 
(\cite{2000AJ....120.1933D}). The XIS FOVs are shown with the solid squares.}\label{fig:rass}
\end{figure*}

\section{Introduction}\label{sec:intro}
The evolution of thermal plasmas in supernova remnants (SNRs) is a dynamical time-sequence of 
shock-heated gas. 
At first, a large fraction of the expanding velocity is converted to random velocity (thermal energy). 
The shock-heated temperature is proportional to the mass of the plasma particles (electrons and ions); 
the ion temperature is more than 1000 times higher than the electron temperature (hereafter, $T_{\rm e}$). 
Subsequently, the energy of ions is transferred to the electrons, and hence $T_{\rm e}$ gradually increases. 
Then the high temperature electrons slowly ionize the ions to higher ionization states. 
Here we define the ionization temperature ($T_{\rm z}$) as a parameter to represent populations of ionization 
states; $T_{\rm z}$ is equal to the plasma temperature in collisional ionization equilibrium (CIE), 
which realizes the observed populations of ionization states. 

The typical timescale of ionization in SNR plasmas (density is $\sim 1$~cm$^{-3}$)  is $\sim 10^{12\--13}$~s 
(\cite{2010ApJ...718..583S}). Accordingly, most of the young-intermediate SNRs have lower $T_{\rm z}$ 
than $T_{\rm e}$, which is called an under-ionized or ionizing plasma, because  ionization dominates 
over recombination. In fact, young--middle-aged shell-like SNRs have been generally described with an 
ionizing plasma model (e.g. \cite{1999PASJ...51..239K}).

Recently, Suzaku has discovered strong radiative recombination continua (RRCs) in the X-ray spectra of 
three SNRs, IC\,443, W49\,B, and G\,359.1$-$0.5 (\cite{2009ApJ...705L...6Y}; \cite{2009ApJ...706L..71O}; 
\cite{2011PASJ...63..527O}). 
The strong RRCs appear when recombination dominates over ionization (recombining plasma). 
In fact, the X-ray spectra of these SNRs can be fit by neither CIE plasma, ionizing plasma, nor any combination 
of them. Detailed spectroscopic analysis with the recombining plasma code revealed that $T_{\rm z}$ are 1.5--3 times 
higher than $T_{\rm e}$. These SNRs have some common features.  (1)~All are categorized into mixed-morphology 
SNRs (MM~SNRs: \cite{1998ApJ...503L.167R}), which have a radio shell with centrally-filled thermal X-rays. 
(2)~These are associated with shocked molecular clouds and GeV/TeV $\gamma$-ray emissions 
(e.g. \cite{2003ApJ...585..319Y};  \cite{2010ApJ...712..459A}). 

W\,28 (G\,6.4$-$0.1), a bright MM~SNR, exhibits above two conditions; a shock interaction with 
the ambient gas indicated by the expanding H\emissiontype{I} shell (\cite{2002AJ....124.2145V}), 
broad CO, CS, and H$_2$ emission lines (\cite{1999PASJ...51L...7A}; \cite{2005ApJ...618..297R}), 
and OH (1720~MHz) masers (\cite{2003ApJ...583..267Y}). Also GeV and TeV $\gamma$-rays were detected (\cite{2010ApJ...718..348A}; \cite{2008A&A...481..401A}). 

From the estimated distance of 1.9$\pm$0.3~kpc (\cite{2002AJ....124.2145V}), 
the angular size of $\sim$\timeform{50'} is $\sim$30~pc. The large size, possible association to 
the nearby pulsar (\cite{1993ApJ...409L..57K}), and dynamical evolution of the H\emissiontype{I} shell 
(\cite{2002AJ....124.2145V}) constrain the age of W\,28 to be 33,000--150,000~yr.

ROSAT and ASCA mapped the entire remnant in the X-ray band (\cite{2002ApJ...575..201R}) and found 
center-filled structure with partial shells at northeast and southwest.  The X-ray spectrum in the central region 
had hard X-ray emissions up to $\sim 7$~keV with the iron (Fe) K$\alpha$ line, and was fit with a two-temperature plasma 
of $0.67$~keV and $1.8$~keV in CIE or ionizing. Thus no hint of recombining plasma has been reported 
from W\,28, although this SNR shares the common features with the three SNRs of recombining plasmas.  The aim of this paper is to 
search for and quantitatively study recombining plasmas with the superior energy resolution and the large effective area of 
the X-ray Imaging Spectrometer (XIS: \cite{2007PASJ...59S..23K}) onboard Suzaku (\cite{2007PASJ...59S...1M}).
Throughout this paper, the distance of 2~kpc is adopted and statistical errors are at the 90\% confidence level. 

\section{Observations and Data Reductions}

We observed the central bright region of W\,28 with the XIS as a part of the recombining plasma survey project 
(PI: Koyama, K.). The local background was obtained from the Suzaku archive near W\,28. The locations of the 
two fields and the observation log are respectively shown in figure~\ref{fig:rass} and in table~\ref{tab:obslog}. 

The XIS consists of four CCD cameras each placed at the focal planes of four X-Ray 
Telescopes (XRTs: \cite{2007PASJ...59S...9S}). Three sensors employ Front-Illuminated (FI) CCDs (XIS\,0, 2, 
and 3), while the other employs a Back-Illuminated (BI) CCD (XIS\,1). 
The entire region of XIS\,2 and one of the edge region of XIS\,0 have not been functional since 
anomalies in 2006 November and in 2009 June, respectively. 
The field of view (FOV) of the XIS combined with XRT covers a \timeform{18'} $\times$ \timeform{18'} region with 
the pixel scale of \timeform{1''}~pixel$^{-1}$. The angular resolution of \timeform{1.9'}--\timeform{2.3'} in the 
half-power diameter is almost independent of photon energies and off-axis angles within $\sim$ 10\%. The total 
effective area of the operational XIS\,0, 1, and 3 combined with the three XRTs is 1070~cm$^2$ at 1.5~keV.
Due to the low orbital altitude of Suzaku at $\sim$550~km, the XIS achieves a low and stable background environment. 

The XIS observation was made with the normal clocking mode.  The software package HEASoft version 6.11 
and the pipeline processing version 2.4 were used for the data reduction. 
To restore the radiation-induced degradation in the energy gain and resolution, the spaced-row charge injection 
technique (SCI: \cite{2004SPIE.5501..111B}) was applied with the makepi files version 20110621 provided by 
the XIS team (\cite{2009PASJ...61S...9U}). Then the systematic uncertainty in the energy scale is $\lesssim$ 
10~eV at 5.9~keV. We removed hot and flickering pixels, and events during the South Atlantic Anomaly passages 
and in the Earth night-time and day-time elevation angles below \timeform{5D} and \timeform{20D}, respectively. 

\section{Analyses and Results}

\begin{figure}[!htbc]
  \begin{center}
    \FigureFile(80mm,90mm){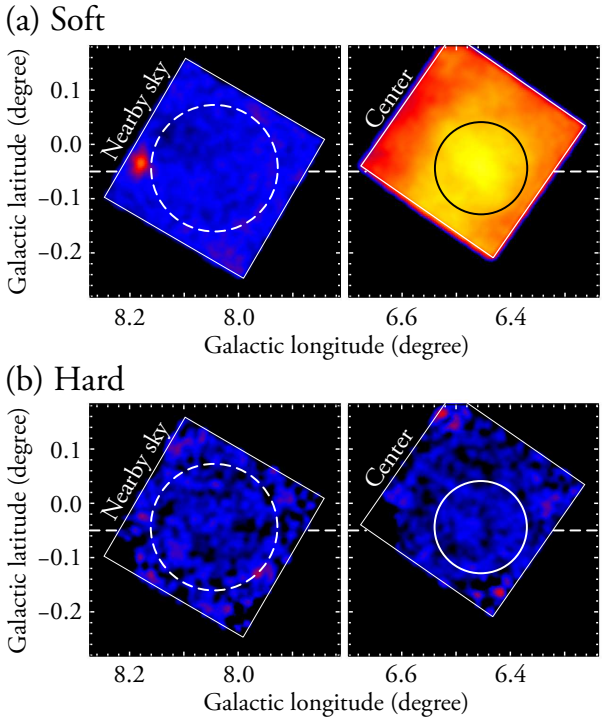}
  \end{center}
  \caption{Band-limited images with the XIS: (a) soft X-rays in 0.5--5.0~keV and 
  (b) hard X-rays in 5.0--8.0~keV. The data with the three CCDs were merged. The NXB 
  was subtracted and then the exposure and vignetting effects were corrected. 
  The source and background extraction regions are shown in the solid and dashed circles, 
  respectively.}\label{fig:xisimg}
\end{figure}
\begin{figure}[!htbc]
  \begin{center}
    \FigureFile(80mm,30mm){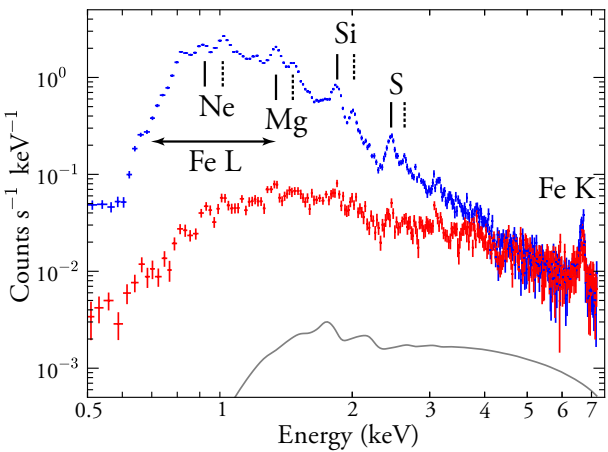}
  \end{center}
  \caption{Comparison of the source (blue) and the normalized background (red) spectra. For visibility, 
only merged spectra of FI (XIS\,0 and XIS\,3) are displayed. The solid and dashed black lines show the 
center energies of the K$\alpha$ lines from He-like and H-like ions, respectively. The gray line indicates the 
CXB spectrum.}\label{fig:3spec}
\end{figure}

\subsection{Background Estimation}\label{sec:bgdsub}
Figure~\ref{fig:xisimg} shows the X-ray images of the W\,28 center and the nearby-sky fields. 
The non--X-ray background (NXB) generated by xisnxbgen (\cite{2008PASJ...60S..11T}) was subtracted, 
and then the vignetting and exposure corrections were made. 

We can see excess emission from W\,28 in the soft X-ray band, but no excess in the hard band (figure~\ref{fig:xisimg}). 
We extracted the source and background spectra from the solid and dashed circles in figure~\ref{fig:xisimg}, respectively. 
The results are shown in figure~\ref{fig:3spec}, where the effective area of the background is normalized to 
that of the source region. 

Compared with the background, the source spectrum shows obvious excess below 
$\sim 5$~keV but no excess in the hard X-ray band  $\gtrsim 5$~keV (figure~\ref{fig:3spec}), 
consistent with the images. This result is, however, inconsistent with the previous ASCA observations, 
which claimed the detection of hard X-ray emission up to $\sim 7$~keV with the Fe K$\alpha$ line 
(\cite{2002ApJ...575..201R}). 

One problem is position-to-position fluctuations of the cosmic X-ray background (CXB). We therefore estimated 
the CXB flux assuming the photon index, the average surface brightness, and the Galactic absorption to be 
1.412, 6.38$\times 10^{-8}$~erg~cm$^{-2}$~s$^{-1}$~sr$^{-1}$ (\cite{2002PASJ...54..327K}), 
and $1.4\times 10^{22}$~H~cm$^{-2}$ (\cite{1990ARA&A..28..215D}), respectively. 
Then the expected CXB level (the gray curve in figure~\ref{fig:3spec}) is only about 5\% of the total background flux. 
Thus the error due to the CXB fluctuation can be neglected. 

The major source of the background is the Galactic diffuse X-ray emission (GDXE) because W\,28 is 
located near the Galactic plane and the Galactic center (GC).  In fact, the Fe K$\alpha$ line at 6.7~keV seen 
in figure~\ref{fig:3spec} is due to the GDXE.  Since the surface brightness of the GDXE quickly decreases 
with the increasing distance from the Galactic plane with the scale height of $\lesssim$ \timeform{1D} 
(\cite{2010PhDT..........U}\footnote{Available at http://repository.tksc.jaxa.jp/pl/dr/IS8000028000/en}), 
we selected the background region from the same Galactic latitude  (figure~\ref{fig:xisimg} left) 
of W\,28 (figure~\ref{fig:xisimg} right). The GDXE uncertainty is, therefore, due to the longitude distribution along the plane. 
The flux of the GDXE decreases as the distance from the GC increases with the scale length  of $\sim$\timeform{30D} 
(\cite{2010PhDT..........U}). Then the GDXE on the background region is estimated to be about 5\% smaller than that 
on W\,28.  Still we found no excess from W\,28 above the background in the hard X-ray band, and hence no 
hard X-ray emission above 5~keV for W\,28 is very conservative conclusion. 

\begin{figure*}[!htbc]
  \begin{center}
    \FigureFile(140mm,220mm){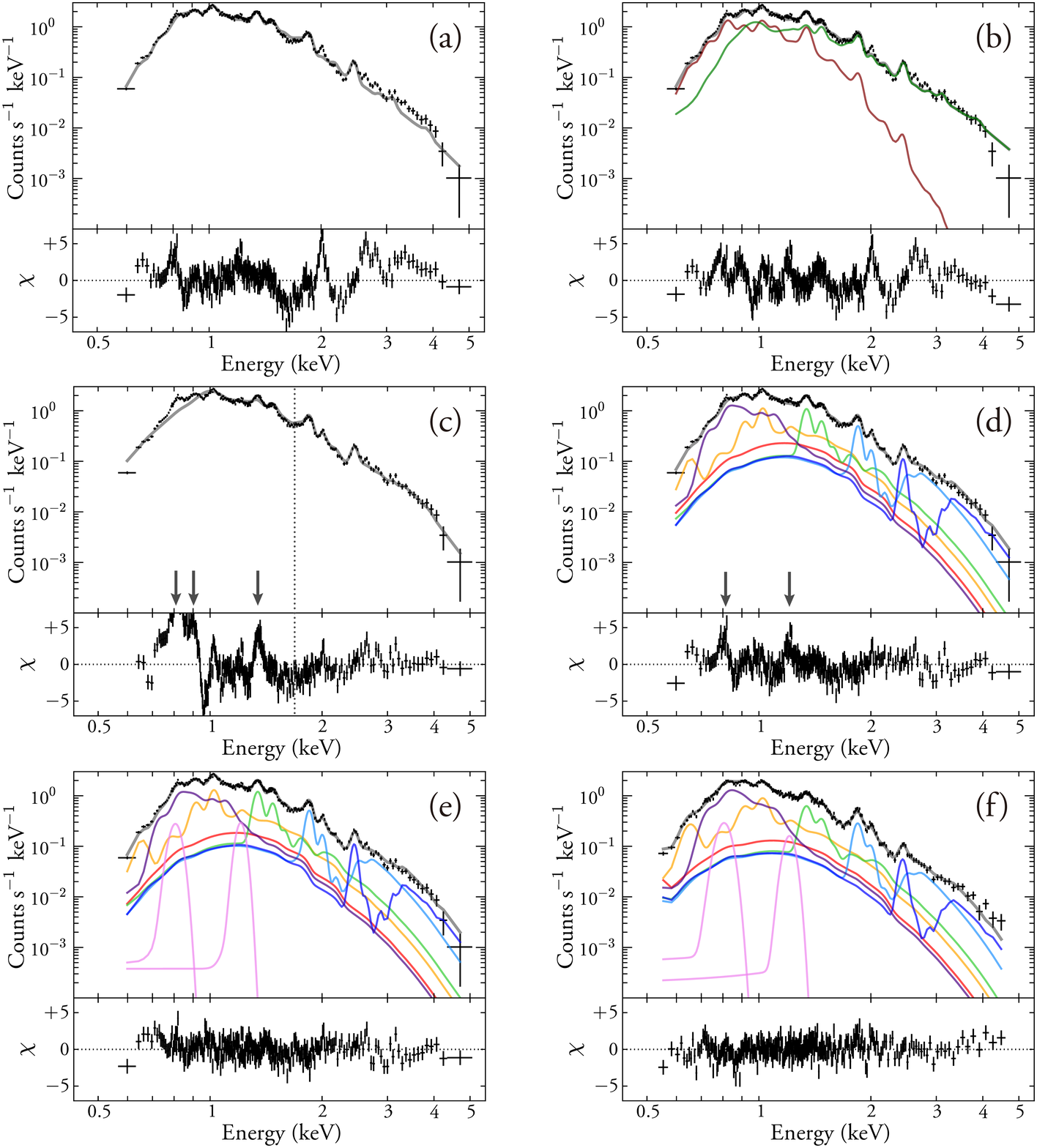}
  \end{center}
  \caption{Background-subtracted spectra (the black crosses) with the best-fit models. 
  For visibility, only data and models for the merged FI are displayed except for (f). Each panel shows results of 
different models: (a) single-temperature CIE, (b) two-temperature CIE, (c) single-$T_{\rm z}$ 
  recombining plasma model, (d) multi-$T_{\rm z}$ recombining plasma model, (e) the same as (d) 
  but artificial Gaussians at 0.8 and 1.2~keV (pink) are added, and (f) the same as (e) but for BI. 
  The thick gray curves are the total emission, while the thin colored curves are each component: 
  low-$kT$ component in dark red and high-$kT$ component in dark green for (b); C and N in red, 
  O and Ne in orange, Mg in light green, Si in light blue, S, Ar, and Ca in dark blue, 
  Fe and Ni in purple for (d), (e), and (f).  
  The solid arrows in (c) indicate the K$\alpha$ line centers of He-like Mg and Ne, and L-shell 
  complex of Fe. The solid arrows in (d) are L-shell transitions of Fe.}\label{fig:spec}
\end{figure*}

\subsection{Spectrum}\label{sec:spec}
We subtracted the background spectrum (the red data in figure~\ref{fig:3spec}) from the source spectrum 
(the blue data in figure~\ref{fig:3spec}). 
The result is shown in figure~\ref{fig:spec}. We can see many K-shell emission lines from highly ionized neon (Ne), 
magnesium (Mg), silicon (Si), and sulfur (S) as well as L-shell emission lines from Fe. We therefore examined 
several optically thin thermal plasma models.  In the spectral analysis, we used the SPEX 
(\cite{1996uxsa.conf..411K}) software version 2.02.02. 
The redistribution matrix and the auxiliary response functions were generated by xisrmfgen and xissimarfgen 
(\cite{2007PASJ...59S.113I}), respectively, with the latest rmfparam files version 20110708. The abundances for 
the absorption of interstellar medium (ISM: \cite{1983ApJ...270..119M}) were assumed to be one solar 
(\cite{1989GeCoA..53..197A}), using the Absm model in SPEX. The abundances for Ne, Mg, Si, S, and Fe in the thermal 
plasma were free parameters, while those of oxygen (O), argon (Ar) and calcium (Ca), and nickel (Ni) were assumed 
to be the same as those of Ne, S, and Fe, respectively. The other elements, including carbon (C) and nitrogen (N) 
were fixed to one solar. 

\begin{table*}[!htbc]
 \caption{Best-fit parameters with CIE and single-$T_{\rm z}$ recombining plasma models.}\label{tab:cieneifit}
 \begin{center}
  \begin{tabular}{lcccccc}
   \hline
   Parameter & Single-temperature CIE & Two-temperature CIE & Single-$T_{\rm z}$ recombining plasma \\
    \hline 
   $N_{\rm H}$ (10$^{21}$~cm$^{-2}$) & 3.90$^{+0.07}_{-0.07}$ & 6.28$^{+0.14}_{-0.15}$ & 2.99 $^{+0.04}_{-0.03}$\\ 
   $VEM_{1}$ (10$^{56}$~cm$^{-3}$)\footnotemark[$*$] & 21.1$^{+0.7}_{-0.6}$ & 103$^{+13}_{-12}$  & 64.6$^{+0.9}_{-0.9}$ \\ 
   $VEM_{2}$ (10$^{56}$~cm$^{-3}$)\footnotemark[$*$] & ... & 13.6$^{+0.5}_{-0.5}$ & ... \\
   $kT_{1}$ (keV) & 0.62$^{+0.01}_{-0.01}$ & 0.24$^{+0.01}_{-0.01}$ & ... \\
   $kT_{2}$ (keV) & ... & 0.77$^{+0.01}_{-0.01}$ & ... \\
   $kT_{\rm e}$ (keV) & ... & ... & 0.43 (fixed) \\ 
   $kT_{\rm z}$ (keV) & ... & ... & 0.95 (fixed) \\ 
   \hline
   Abundance (solar) \\
   ... Ne & 0.39$^{+0.02}_{-0.02}$ & 0.13$^{+0.01}_{-0.01}$ & $\le 0.01$ \\ 
   ... Mg & 0.36$^{+0.02}_{-0.02}$ & 0.41$^{+0.02}_{-0.02}$ & 0.18$^{+0.01}_{-0.01}$ \\ 
   ... Si  & 0.25$^{+0.01}_{-0.01}$ & 0.32$^{+0.02}_{-0.02}$  & 0.20$^{+0.01}_{-0.01}$ \\ 
   ... S & 0.50$^{+0.04}_{-0.04}$ & 0.43$^{+0.04}_{-0.03}$ & 0.26$^{+0.02}_{-0.02}$ \\ 
   ... Fe & 0.14$^{+0.01}_{-0.01}$ & 0.23$^{+0.01}_{-0.01}$ & 0.06$^{+0.01}_{-0.01}$ \\ 
   \hline
   $\chi^2/$d.o.f. & 2160/641 & 1753/639 & 6914/642 \\ 
   \hline
   \multicolumn{3}{@{}l@{}}{\hbox to 0pt{\parbox{170mm}{\footnotesize
   \par\noindent
   \footnotemark[$*$] Volume emission measure at the distance of 2~kpc: 
   $VEM = \int n_{\rm e}n_{\rm p} dV$, where $n_{\rm e}$, $n_{\rm p}$, and $V$, are 
   the electron and proton densities, and the emitting volume, respectively.
   }\hss}}
  \end{tabular}
 \end{center}
\end{table*}

\subsubsection{CIE Plasmas}
We tried a single-temperature optically thin thermal plasma model in CIE  (the Cie model in SPEX). 
The result is shown in figure~\ref{fig:spec}a. Although we fit simultaneously for the FI and BI spectra, we show 
only the best-fit FI spectrum for simplicity. As is seen in figure~\ref{fig:spec}a, this model obviously failed to reproduce 
the spectrum with $\chi^2/$d.o.f. of 2160/641. 
We therefore examined a two-temperature CIE model (figure~\ref{fig:spec}b), with coupled abundances 
for each element between the two CIE components. This model also failed to reproduce the spectrum with 
$\chi^2/$d.o.f. of 1753/639. The best-fit parameters for these CIE models are given in table~\ref{tab:cieneifit}. 
In addition to the 0.77-keV plasma, W\,28 requires 0.24-keV plasma instead of the 1.8-keV plasma 
of the ASCA results (\cite{2002ApJ...575..201R}). The authors proposed that W\,28 is a 
peculiar MM~SNR because of the presence of the high-temperature plasma of 1.8~keV. 
Our results of the 0.24+0.77-keV temperature plasma revised the interpretation of W\,28 
to be a typical MM~SNR (0.4--0.9-keV temperature; \cite{1998ApJ...503L.167R}). 
This revised picture comes from the absence of the high energy photons above 
$\sim 5$~keV in the Suzaku spectrum, which would be due to proper background subtraction 
(subsection~\ref{sec:bgdsub}). To improve the value of $\chi^2/$d.o.f., we added further CIE components, 
but found no significant improvement. 
The most remarkable residuals in any CIE model are line-like excess at the Si Lyman$\alpha$ energy (2.0~keV) 
and bump-like structures at 2.4--5.0~keV.

\subsubsection{Recombining Plasma Model with Single $T_{\rm z}$}
The bumps at  2.4--5.0~keV have two rising edges at 2.4 and 3.2~keV, which are 
most probably due to RRCs of He-like Si and S, respectively (\cite{2011PASJ...63..527O}).  
Together with the strong Si Lyman$\alpha$ residual, 
the spectrum is due to a recombining plasma, at least for Si and S. We thus tried a recombining plasma model 
in the RRC band of Si and S (1.7--5.0~keV).  Then the residuals found in the 1.7--5.0~keV band disappear with 
the best-fit $T_{\rm z}$ of $0.95^{+0.01}_{-0.09}$~keV and $T_{\rm e}$ of $0.43^{+0.03}_{-0.01}$~keV. 
Fixing these best-fit temperatures, we made a full energy band fit. The free parameters were abundances of 
Ne, Mg, Si, S, and Fe, and $N_{\rm H}$. The result is, however, unacceptable with 
$\chi^2/$d.o.f. of 6914/642 (figure~\ref{fig:spec}c). As shown by the solid arrows in figure~\ref{fig:spec}c, 
remarkable excess are found at the K$\alpha$ lines of He-like Mg (1.35~keV)  and Ne (0.92~keV) and the Fe L-shell 
complex at $\sim$0.8~keV. 

\subsubsection{Multi-$T_{\rm z}$ Recombining Plasma}
The data excess near the K$\alpha$ lines of He-like Mg and Ne 
indicates underestimation of the He-like line fluxes compared to the H-like line 
(see the solid arrows in figure~\ref{fig:spec}c). 
This supports that $T_{\rm z}$ for Mg is significantly lower than those of Si and S. 
We therefore introduced a recombining plasma model with different $T_{\rm z}$ for each element (figure~\ref{fig:spec}d). 
Since no spectral code to directly describe such a plasma is available, we approximated the model  with the combination 
of several single-$T_{\rm z}$ plasmas whose $T_{\rm e}$ are coupled among the all plasmas. We grouped the elements 
into six sub-groups assuming the same abundances in each sub-group (see the caption of figure~\ref{fig:spec}). 
For C and N, we fixed $T_{\rm z}$ to $T_{\rm e}$. 

\begin{table}[!htbc]
 \caption{Best-fit parameters with a multi-$T_{\rm z}$ recombining plasma model.}\label{tab:mtzfit}
 \begin{center}
  \begin{tabular}{lcccccc}
   \hline
   Parameter & A & B \\ 
    \hline 
   $N_{\rm H}$ (10$^{21}$~cm$^{-2}$) & 4.65$^{+0.15}_{-0.12}$ & 4.70$^{+0.15}_{-0.11}$ \\ 
   $VEM$ (10$^{56}$~cm$^{-3}$)\footnotemark[$*$] & 60.7$^{+9.3}_{-7.1}$ & 47.1$^{+7.3}_{-5.8}$ \\ 
   $kT_{\rm e}$ (keV) & 0.39$^{+0.02}_{-0.02}$ & 0.40$^{+0.02}_{-0.03}$ \\ 
   \hline
   $kT_{\rm z}$ (keV) \\
   ... Ne & 0.48$^{+0.04}_{-0.03}$ & 0.46$^{+0.05}_{-0.04}$ \\ 
   ... Mg & 0.66$^{+0.06}_{-0.05}$ & 0.66$^{+0.07}_{-0.06}$ \\ 
   ... Si  & 1.01$^{+0.07}_{-0.06}$ & 1.01$^{+0.08}_{-0.08}$ \\ 
   ... S & 0.95$^{+0.11}_{-0.09}$ & 0.96$^{+0.14}_{-0.10}$ \\ 
   ... Fe & 0.58$^{+0.04}_{-0.04}$ & 0.58$^{+0.05}_{-0.05}$ \\ 
   \hline
   Abundance (solar) \\
   ... Ne & 0.16$^{+0.01}_{-0.02}$ & 0.23$^{+0.03}_{-0.02}$ \\ 
   ... Mg & 0.32$^{+0.02}_{-0.02}$ & 0.44$^{+0.04}_{-0.03}$ \\ 
   ... Si  & 0.26$^{+0.02}_{-0.03}$ & 0.34$^{+0.03}_{-0.04}$ \\ 
   ... S & 0.33$^{+0.08}_{-0.08}$ & 0.40$^{+0.09}_{-0.09}$ \\ 
   ... Fe & 0.10$^{+0.01}_{-0.01}$ & 0.11$^{+0.01}_{-0.01}$ \\ 
   \hline
   $\chi^2/$d.o.f. & 1268/636 & 893/634 \\
   \hline
   \multicolumn{3}{@{}l@{}}{\hbox to 0pt{\parbox{70mm}{\footnotesize
   \par\noindent
   \footnotemark[$*$] The value at the distance of 2~kpc.
   }\hss}}
  \end{tabular}
 \end{center}
\end{table}

As shown in figure~\ref{fig:spec}d, this model (model\,A) well reproduced the overall structure of the 
observed spectrum. Still some residuals remain at $\sim 0.8$ and $\sim 1.2$~keV 
(see the solid arrows in figure~\ref{fig:spec}d). These features 
are known to originate from incomplete atomic data in the current spectral code. 
The former comes from the uncertain intensity ratio of the Fe L-shell
transitions of 3s$\rightarrow$2p and 3d$\rightarrow$2p (\cite{2007ApJ...670.1504G}
and reference therein). The present code may underestimate
the 0.8~keV line flux.
The latter is due to the lack of a number of the Fe L-shell lines by the 
transitions from highly excited states of $n\ge5$ (\cite{2000ApJ...530..387B}). 
Thus, we artificially added two Gaussians at 0.8~keV and 1.2~keV (figure~\ref{fig:spec}e and~\ref{fig:spec}f). 
Then the fit largely improved (model\,B; $\chi^2/$d.o.f. = 893/634).  The best-fit parameters for 
Model\,A and B are listed in table~\ref{tab:mtzfit}. 
We see that the best-fit $T_{\rm z}$ are essentially the same between  model\,A and~B. The best-fit $T_{\rm z}$ 
are different among elements with higher temperature than $T_{\rm e}$ (table~\ref{tab:mtzfit}). 

\begin{figure*}[!htbc]
  \begin{center}
    \FigureFile(170mm,80mm){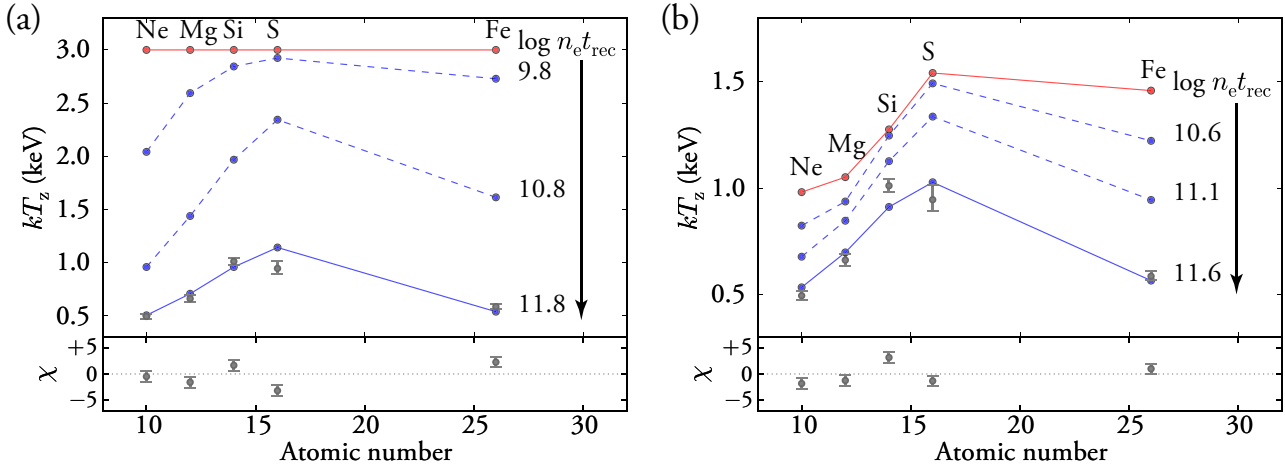}
  \end{center}
  \caption{Time evolution  of $T_{\rm z}$ in recombining plasmas with the initial conditions of (a) $T_{\rm z0} = 3.0$~keV for all elements,  and  (b) $T_{\rm z0}$ realized by an ionizing plasma with $T_{\rm e}' = 10$~keV 
  and $n_{\rm e}t_{\rm ioni} = 10^{11}$~s~cm$^{-3}$. The gray circles with bars 
  show the observed  $T_{\rm z}$ with 1-$\sigma$ statistical errors. 
  The direction of evolution as a function of $n_{\rm e}t_{\rm rec}$ is indicated by the vertical arrows. 
  The solid lines in red and blue respectively show the initial and final states, while the dashed blue lines 
  show transitional states. 
  }\label{fig:plasim}
\end{figure*}

\section{Discussion}

We have found that the observed spectrum can be reproduced by a multi-$T_{\rm z}$ recombining plasma. 
The recombining plasmas in SNRs are not expected in the standard scenario, and hence there must be a 
missing branch of evolution of SNRs. Several ideas have been proposed, such as rapid electron cooling 
by adiabatic rarefaction (\cite{1989MNRAS.236..885I}), or conduction cooling by ambient cold matter 
(\cite{2002ApJ...572..897K}), extra-ionization by high-energy photons (\cite{2002ApJ...572..897K}) or 
by supra-thermal particles (\cite{2011PASJ...63..527O}), and so on.  Any scenario may compose of two
processes;  (1) initial process to make a plasma of higher $T_{\rm z}$ than  $T_{\rm e}$ , and (2) gradual relaxation of the recombining plasma toward CIE.  
We propose that the process (2) carries the essential roll to make element-dependent $T_{\rm z}$. Based 
on the simulation of the $T_{\rm z}$ relaxation in time and comparison with the observed multi-$T_{\rm z}$ 
values, possible origin of the recombining plasma from W\,28 is discussed in the following subsections.  

\subsection{Recombination Timescale}\label{sec:rec}
Difference of $T_{\rm z}$ among elements can appear in a relaxation epoch of recombining  plasma to 
CIE because the relaxation timescale is different from element to element. 
Numerical studies have demonstrated that the relaxation timescale basically shows monotonic 
increase with atomic number, except for Fe and Ni (e.g., see figure~1 of \cite{2010ApJ...718..583S}). 
These two heavy elements have shorter timescales than the other lighter elements 
at $T_{\rm z}$ from sub to a few~keV,  because the L-shell processes are dominant for Fe and Ni, 
while the K-shell processes are dominant for the others in this temperature range. 
The observed $T_{\rm z}$ values appear to satisfy this trend (table~\ref{tab:mtzfit}). 

To examine whether the difference in recombination timescale can quantitatively explain the observed 
$T_{\rm z}$, we simulate time evolutions of $T_{\rm z}$ in a recombining plasma by using the Neij model in 
SPEX. The parameters are the initial ionization temperature $T_{\rm z0}$, $T_{\rm e}$, and the recombination 
time $n_{\rm e}t_{\rm rec}$, where $n_{\rm e}$ and $t_{\rm rec}$ are the density of electrons and the elapsed 
time, respectively.  

We have tried many sets of ($T_{\rm z0}$, $t_{\rm rec}$) for fixed $T_{\rm e}$ of 0.4~keV following 
the best-fit value (table~\ref{tab:mtzfit}). We obtained similar best-fit $T_{\rm z}$ distribution from 
different parameter set of  ($T_{\rm z0}$, $t_{\rm rec}$). We therefore constrain the parameter set within 
physically reasonable values.  One plausible case of $T_{\rm z0}=3$~keV is given in figure~\ref{fig:plasim}a, 
which shows the simulated decrease of $T_{\rm z}$ with time ($n_{\rm e}t_{\rm rec}$). We find the best-fit 
recombination time to be $n_{\rm e}t_{\rm rec} = 10^{11.8}$~s~cm$^{-3}$. The simulated $T_{\rm z}$ values 
after $10^{11.8}$~s~cm$^{-3}$, is close to the observation with $\chi^2/$d.o.f. of $\sim 21/4$.  We caution 
that the $\chi^2/$d.o.f. value is just for reference to compare a more complex simulation given in the 
next subsection, and should not be taken seriously, because the major uncertainty of the observed $T_{\rm z}$ 
is not statistic error but systematic one due to incomplete modeling and/or atomic data in the analyses. 

\subsection{Initial Variation of $T_{\rm z}$}\label{sec:init}
In a more general case, initial $T_{\rm z0}$ should be different among elements. 
We hence simulate the time evolution of $T_{\rm z}$ in this case. The simulation consists of two steps. 
In the first step, we simulate the evolution of $T_{\rm z}$ in an ionizing plasma. 
This gives the element-dependent $T_{\rm z0}$. Then using this result, we simulate a recombining 
plasma as was done in the previous subsection. 

The parameters are the electron temperature during the ionizing phase $T_{\rm e}'$, the ionization time 
$n_{\rm e}t_{\rm ioni}$, and the recombination time $n_{\rm e}t_{\rm rec}$.  The electron temperatures 
during the two processes, $T_{\rm e}'$ and $T_{\rm e}$, are assumed to be 10~keV and 0.4~keV, 
respectively. 

In figure~\ref{fig:plasim}b, we plot decrease of $T_{\rm z}$ during the recombining 
phase for the case of $T_{\rm e}' = 10$~keV and $n_{\rm e}t_{\rm ioni} = 
10^{11.0}$~s~cm$^{-3}$.  This combined process gives a better fit to the observed $T_{\rm z}$, 
than that of the single relaxation process of recombining plasma (see subsection~\ref{sec:rec}) at 
$n_{\rm e}t_{\rm rec} = 10^{11.6}$~s~cm$^{-3}$ with $\chi^2/$d.o.f. of $\sim 17/4$. 
One may argue that other combinations of the parameter values may also work as well. As we already noted, 
many combinations were examined and the recombination times are always found to be 
$\sim$10$^{11.5}$~s~cm$^{-3}$ or longer.  Thus, the result of $n_{\rm e}t_{\rm rec}$ does not 
depend much on the initial conditions of $T_{\rm z0}$. 

\subsection{Origin of Recombining Plasma in W\,28}

We have shown that the combination of the processes (1) and (2) provides a plausible scenario for producing the observed spectrum with the initial conditions given in figure 5(b) (see the beginning of Discussion and section 4.1). The initial conditions can be set by collisional ionization in thermal pool of electrons with no special ionization  process such as photo-ionization or supra-thermal electrons. 
This supports the electron cooling scenario for W\,28. In the simulation, $n_{\rm e}t_{\rm rec}$ is 
$\sim 10^{11.6}$, which corresponds to $\sim 10^{4} \times (n_{\rm e}/{\rm 1\:cm}^{-3})^{-1}$~yr, roughly 
consistent with the age of W\,28. Thus, the recombination time is consistent with 
the rarefaction scenario (\cite{1989MNRAS.236..885I}). 

This recombination time is also nearly the same order of the cooling timescale by 
thermal conduction, which is estimated to be $\sim$2$\times$10$^{4} \times (n_{\rm e}/{\rm 1\:cm}^{-3})$~yr 
for the plasma of 6-pc diameter (see equation~5 in \cite{2002ApJ...572..897K}). However, we found 
no spatial correlation of the recombining plasma to the molecular clouds (Sawada et al. 2012, in preparation), 
which may not favor the thermal conduction scenario. We therefore propose the rarefaction scenario 
(\cite{1989MNRAS.236..885I}), and discuss along this scenario in the following paragraphs. 
More quantitative comparison of these two models will be given in the next paper using the results of 
spatial distribution of the recombining plasma (Sawada et al. 2012, in preparation). 

The initial event of the rarefaction scenario would be a type-II supernova explosion in 
dense circum-stellar medium (CSM) made by stellar winds from the massive progenitor. 
Assuming a spherical plasma with the diameter of 6~pc, the mass of the X-ray emitting gas 
in the center region is estimated to be $\sim$5~$M_{\odot}$. This value is similar to that 
of a massive progenitor, and hence provides further circumstantial support for the 
rarefaction scenario. 
The shock-heated ejecta and CSM experience rapid electron cooling due to adiabatic expansion after 
shock break-out to the rarefied ambient ISM. 
The break-out and hence the cooling, occur at the early phase of 
the SNR evolution (about several 10--100~yr after the explosion; \cite{1989MNRAS.236..885I}). 
Thus most of the life would be in the relaxation phase of the recombining plasma, consistent with 
the large $n_{\rm e}t_{\rm rec}$ values obtained in our simulations. 
Due to the dense environment of CSM, the shocked gas quickly reaches a temperature of several~keV 
before the break-out,  in good agreement with the $T_{\rm z0}$ values assumed in our simulations. 

\citet{2011arXiv1110.4948S} extended the previous calculation of \citet{1989MNRAS.236..885I} to 
non-symmetric cases and found that center-filled X-ray morphologies are realized with anisotropic CSM 
due to the rotation of a progenitor star. The size of the X-ray emitting plasma is about a few 10~pc, which is 
consistent with that of W\,28.  Therefore, this scenario can explain both the spectral and spatial structures 
of the X-ray emission from W\,28. 

\bigskip
The authors thank K.~Masai, T.~Shimizu, and H.~Yamaguchi for valuable discussion, 
J.~Kaastra and J.~de~Plaa for their useful comments on spectral modeling, and 
S.~Nakashima, M.~Nobukawa, H.~Uchida, and H.~Uchiyama for improving the draft. 
We also thank all of the Suzaku team members for their full support of the Suzaku project. 
We acknowledge the financial support from the Ministry of Education, Culture, Sports, Science 
and Technology (MEXT) of Japan; the Grant-in-Aid for the Global COE Program ``The Next 
Generation of Physics, Spun from Universality and Emergence'' and others for Scientific Research B 
(No. 20340043 and 23340047).
MS is supported by Japan Society for the Promotion of Science (JSPS) Research Fellowship for Young Scientists.
KK is supported by the Challenging Exploratory Research program.

\end{document}